# HistoColAi: An Open-Source Web Platform for Collaborative Digital Histology Image Annotation with AI-Driven Predictive Integration

Cristian Camilo Pulgarín-Ospina[a], Rocío del Amor[a], Adrián Colomer[a,b], Julio Silva-Rodríguez[c] and Valery Naranjo[a,b]

[a]Instituto Universitario de Investigación en Tecnología Centrada en el Ser Humano, Universitat Politècnica de València, Valencia, Spain
[b]valgrAI: Valencian Graduate School and Research Network of Artificial Intelligence, Valencia, Spain
[c]ÉTS Montréal, Montréal, Québec, Canada,

ABSTRACT

Digital pathology has become a standard in the pathology workflow due to its many benefits. These include the level of detail of the whole slide images generated and the potential immediate sharing of cases between hospitals. Recent advances in deep learning-based methods for image analysis make them of potential aid in digital pathology. However, a major limitation in developing computer-aided diagnostic systems for pathology is the lack of an intuitive and open web application for data annotation. This paper proposes a web service that efficiently provides a tool to visualize and annotate digitized histological images. In addition, to show and validate the tool, in this paper we include a use case centered on the diagnosis of spindle cell skin neoplasm for multiple annotators. A usability study of the tool is also presented, showing the feasibility of the developed tool.

## 1. Introduction

A routine (non-digital) pathology workflow typically involves manual steps such as tissue procurement and processing, slide creation, and microscopic analysis. In the best scenarios, a hospital has at least an expert pathologist for slide interpretation or a nearby hospital with a pathology group to assist this process. However, it may not be the regular context. Consequently, the slides are transferred and sent to faraway pathologists, making the workflow laborious, time-consuming, and with more stages [1].

The deployment of whole slide imaging (WSI) technology has attracted significant attention in the pathology community, where the use of digital pathology (DP) for primary diagnosis is becoming commonplace [2]. Digital pathology allows the scanning of entire glass slides, providing diagnostic-quality digital images (WSIs). The high-quality WSIs enable powerful zooms without losing the image definition, aiding the diagnostic process. Additionally, using an image viewer, WSIs can be shared with multiple pathologists worldwide through the internet without needing a microscope, eliminating delays in sample shipments. For all these reasons, DP has the potential to transform the way pathology services are delivered, providing a flexible platform to improve the safety, quality, and efficiency of Pathological diagnosis while preparing for the future of an increasingly pressured medical specialty [3].

Recent advances in deep learning (DL)-based methods for image analysis provide a potential aid in digital pathology. The deployment of convolutional neural networks (CNNs) has allowed the automatic identification of new biomarkers and innovative features in different imaging modalities that support the diagnostic process [4]. DL methodologies have had successful results when applied to pathological imaging with different cancers, i.e., colorectal cancer ([5]; [6]; [7]; [8]; [9]), prostate cancer ([10]; [11]; [12]; [13]) and malignant melanoma ([14]; [15]; [16]; [17]; [18]) among others.

DL models should be trained on large datasets with annotated heterogeneous tissue structures to achieve optimal performance. However, this annotation is difficult in many cases, making it impossible to develop robust models that can be used in practice. One of the main reasons for the difficulty in obtaining massive annotated histopathological datasets is the absence of an open source web-based tool to facilitate the annotation task for pathologists. *HistoColAi*, comes to solve this limitation. In order to demonstrate the advantages of the proposed annotation tool, in this work we present firstly the comparison with the tools that already exist in the scientific community. In the next sessions of the paper, we describe the application structure and its validation by means of a crowdsourcing experiment and a usability study.

### 1.1. Comparison of annotation tools

We have compared our developed tool with seven applications: imageJ ([19]), Cell Profilers ([20]; [21]), Icy ([22]; [23]), Fiji ([24]), Qupath ([25]; [26]; [27]), Ilastik ([28]; [29]), and Upath ([30]). The comparison, as can been observed in Table 1, has been carried out regarding different aspects:

- Accessibility: *HistoColAi* is a multiplatform application accessible from any web browser being this one of the main advantages of the software. Only ImageJ and Upath include a web service.

- Availability: *HistoColAi* is an open source tool. All the software's included in the comparison are also open source except Upath.

ORCID(s): 0000−0002−9759−3791 (C.C. Pulgarín-Ospina)

- WSI management: Our application has been optimized to offer the best experience in terms of usability and performance in the management of WSIs, therefore it is an advantage over other generalist tools, which are not developed ad-hoc to work with this type of image. Upath, and Qupath are the only tools that work with WSIs. The rest of the compared tools are not optimized to work with this image format.

- Usability: Our web application has been developed with a focus on the pathologist's work, aiming to help them to speed up their work. To validate the developed application and demonstrate its usability, we present in this paper a crowdsourcing experiment with 10 non-expert pathologists. As will be presented in the results, the proposed tool significantly reduces the annotation time from hours to a few minutes, while maintaining high fidelity with the manual annotations made by expert pathologists. Most of the compared tools are not user-friendly and have more complex functionalities. Only Qupath seems to be designed with the pathologist's work in mind, but even so, the functionalities provided may be more challenging for inexperienced users.

- Model embedding: (*HistoColAi*) is designed to incorporate already developed models. The use case presented in this paper incorporates a deep learning-based model to automatically detect tumour regions in WSI. These automatic predictions help pathologists to increase the efficiency of WSI analysis.

- Crowdsourcing annotation: *HistoColAi* includes a crowdsourcing system that enables knowledge sharing, allowing the labeling of confusing regions of interest by more than one pathologist. None of the applications included in the comparison offer this functionality.

None of the solutions analyzed previously before the development of *HistoColAi* met all the necessary requirements to be used free of charge in the development and validation of artificial intelligence models on WSI images. This makes necessary the development of a new tool, *HistoColAi*, whose features make it an indispensable tool for the aforementioned objective as well as a potential clinical tool to speed up the diagnosis and annotation process.

The main contributions presented in this work are:

- The architecture and functionalities of a novel open source web application for the visualisation and annotation of WSI. The tool also allows the embedding of artificial intelligence models and collaborative annotation, as well as the control of users' annotation times.

- The results of the validation of this tool through the development of a use case in a real scenario with 10 pathologists in training:

  – An automatic tumour region detection model based on convolutional networks.
  – The results of the tool usability assessment.

The code and trained models are publicly available at https://github.com/cripulos/HistoColAi.

## 2. Materials

To validate *HistoColAi*, we use the database *CR-AI4SkIN* dataset. This database is composed of two private databases (DSV and DSG) from the University Clinic Hospital of Valencia and San Cecilio University Hospital in Granada (Spain). DSV and DSG are composed of histopathological skin images from different body areas that contain cutaneous spindle cell (CSC) neoplasms, i.e, leiomyomas (lm), leiomyosarcomas (lms), dermatofibromas (df), dermatofibrosarcomas (dfs), spindle cell melanomas (mfc), fibroxanthomas (fxa) and squamous cell carcinoma (cef). Each database (DSV and DSG) comprises 180 and 91 different patients, respectively, who signed the pertinent informed consent. Therefore, the complete database (*CR-AI4SkIN*) dataset contains 271 WSIs. A summary of the database description is presented in Table 2.

## 3. Methodology

In this section we present the tool and the methodology followed for its validation. As mentioned above, a use case for the detection of skin cancer tumoral regions is presented, with the aim of validating the proposed system. Although this use case will be described throughout the paper, Figure 1 provides an overview of the workflow to be followed for using *HistoColAiP* in a typical case of segmentation model development.The following subsections explain the architecture of the developed platform and the interconnection of the different modules.

### 3.1. HistoColAi Description

#### 3.1.1. Infrastructure

hanks to advances in web technologies and communication protocols, we have moved from simple web services with static content to complex systems that allow users to share large documents or view high-quality multimedia content via streaming. Taking advantage of the benefits and technologies offered by the web, we propose an innovative application for producing and managing gigapixel images in WSI format in digital pathology.

**Service environment.**

*HistoColAi* is a web service accessible via browser from any device with internet access. The infrastructure of the application consists of two parts: Frontend,the part that runs on the client side, and Backend the part that runs over server side. See Figure 2.

| | ImageJ(19; 31) | CellProfiler(20; 21) | Icy(22; 23) | Fiji(24; 32) | QuPath(26; 27) | Ilastik(28; 29) | Upath(30) | HistoColAi |
|---|---|---|---|---|---|---|---|---|
| Accessibility | Y | N | N | N | N | N | Y | Y |
| Availability | Y | Y | Y | Y | Y | Y | N | Y |
| WSI managemet | N | N | N | N | Y | N | Y | Y |
| Usability | N | N | N | N | N | N | N | Y |
| Model embedding | N | N | N | N | N | N | Y | Y |
| Crowdsourcing annotation | N | N | N | N | N | N | N | Y |

**Table 1**
Comparison of tools for the visualisation and analysis of histopathological images. Red "N" indicates that the software does not have the analysed functionality, green "S" indicates that it has the functionality.

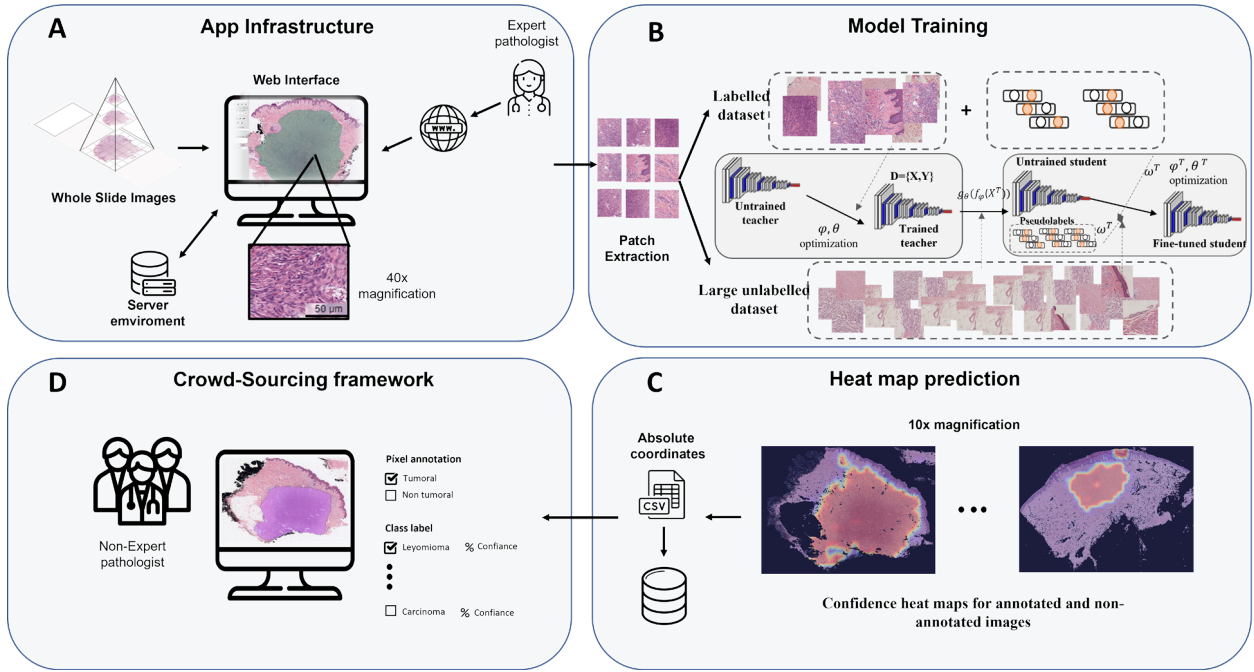

**Figure 1:** Overview of the use case workflow. **A** Each WSI is converted to "dzi" format and uploaded to the system for visualisation in a web environment. Once uploaded, expert pathologists annotate regions of interest (tumoral regions) on a few images. **B** The images annotated by the experts are used to train the deep learning model and automatically predict the tumoral regions of all WSIs (master-student approach). Note that the model is trained at patch-level due to the large size of the WSIs. **C** The patch-level predictions are combined to obtain a prediction map (heat map) over the entire WSI. These maps are thresholded to obtain masks of tumoral regions that are fed into the system. **D** These masks, obtained by the model, are validated by a set of pathologists in training.

**Table 2**
Database distribution. DSV: database from Valencia; DSG: database from Granada. Lm:leiomyomas; lms: leiomyosarcomas; df:dermatofibromas; dfs: dermatofibrosarcomas; fxa: fibroxanthomas; spindle cell melanomas; cef: squamous cell carcinoma.

| | lm | lms | df | dfs | mfc | fxa | cef | Total |
|---|---|---|---|---|---|---|---|---|
| DSV | 28 | 19 | 52 | 11 | 32 | 28 | 10 | 180 |
| DSG | 27 | 9 | 16 | 7 | 6 | 26 | - | 91 |
| Total | 55 | 28 | 68 | 28 | 38 | 44 | 10 | 271 |

The front of web services consists mainly of two technologies: HTML (Hypertext Markup Language), which allows defining of basic web structures and objects, and CSS (Cascading Style Sheets) language, which gives visual richness to the objects and structure defined with HTML. Using

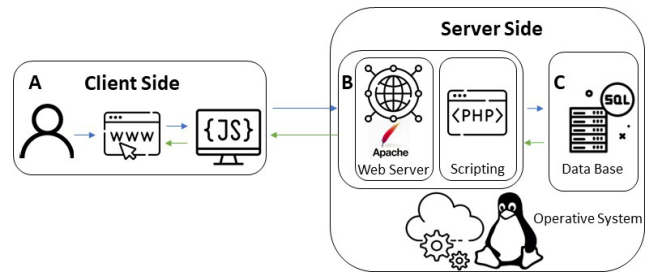

**Figure 2:** Application stack overview is divided into two parts: frontend (Client Side, **A** box); this part runs over the browser as a web service and is coded in HTML, CSS, and JS. The backend (Server Side, **B** and **C** box) runs over the server and has a LAMP stack; Linux as OS, Apache as a web server, PHP backend language, MySQL for a database.

HTML and CSS results in a static web with non interacting content, which is why today's web pages use JavaScript (JS). JS allows the implementation of complex functions executed on the client side (see figure 2.A), relieving the server of the execution burden. In addition, its use allows the integration of libraries to manage complex objects such as drawings and shapes on a web canvas, enabling the interaction of objects with the user.

On the backend, a "LAMPA" stack (Linux, Apache, PHP, and MySQL) is used; Linux as the server operating system, Apache as a web server, PHP (see Figure 2.B) as web server scripting language, MySQL as database engine (see Figure 2.C). The PHP scripting language enables backend communication. JavaScript and MySQL, thus allowing the request and response of data between client-server-database (see Figure 2).

**Data upload.** Web management of heavy multimedia formats is not trivial, as the web is designed to load light data that does not overload a website's reading and rendering time. In order to meet this requirement, protocols have been created that segment multimedia content into small, lightweight fragments, thus achieving optimal web load times. This file size problem is something we have to face with the *HistoColAi* application. To overcome this problem, we use *Pyvip* (33). This library allows us to crop large images, such as WSI images, resulting in a folder system representative of the whole image. Each subfolder stores the number of tiles into which the WSI is divided according to the zoom level at which the image is viewed; see Figure 3. This Pyvip-generated file system allows WSIs to be loaded into the application frontend rendering at different zoom levels suitable for loading and viewing in a browser. Thanks to this library, it is possible to split the WSI of a scanned biopsy at various zoom levels and display it easily and fluently in a web service.

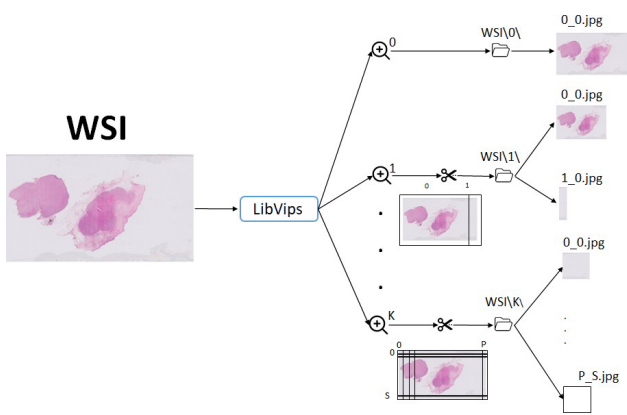

**Figure 3:** WSI batch processing. WSI is processed with *LibVips* (33). As a result, $K$ subfolders are generated containing the slices that compose the WSI at each zoom level. These images are lightweight, so they can be viewed in a web browser.

**Data visualization.** A highlight of the developed application is the ability to annotate WSI sets in a web service, which involves manipulating and rendering gigapixel images in a browser. To achieve the loading and display of WSIs without loss of quality, *HistoColAi* uses the *Openseadragon* Javascript library (34). With this library, it is possible to load the necessary data to visualize the part of the tissue that the user wants to see according to the zoom level and the part of the WSI that is being analyzed.

To access the app, users must use a URL where the set of images or resources previously uploaded to be displayed is indicated; this set of WSI or resources is called a batch. Does the access URL have the following structure: `http://Server/web_folder/index.html?source=images/batch/dzi_images.json`. The first part of the URL goes all the way to the question mark `http://Server/web_folder/index.html?`, indicates the server or domain name address, location and resource (HTML and CSS) to be processed by the client browser, after which the browser loads the application structure and formatting. The second part `source=images/batch/dziimages.json` specifies the location of the batch. The batch comprises several WSIs where the path to the folders representing each WSI is defined in the loaded JSON file. Each path points to a folder containing the WSI fragment at different zoom levels. The methods implemented in the *Openseadragon* library allow using APIs to communicate with the server to get the image to print on the canvas. Therefore, when a client uses the URL, the client side loads the file 'dziimages.json", which contains the paths to all the slide images in the batch, and the frontend makes a GET API with the path of the slide in order to obtain the tissue images to display on the browser.

**Data annotation.** The proposed software allows the annotation of tumor regions and other structures in WSIs using the web service. In order to perform the segmentation and annotation work, a set of tools are needed to perform this task. To provide these functionalities, we use the *paper.js* library (35), which allows the creation of graphical objects and freehand segmentation with different colors, or contours. Figure 4 shows the tools provided for the annotation process. The main functionalities of each module are detailed below:

- Figure 4 A) shows the display and navigation module. This module consists of a navigation box indicating the section of the image displayed in the main view (see Figure 4 E), navigation tools to scroll the image display area and change the zoom level, and arrows to return to the previous image or continue to the next one, and navigate through the different images. The navigation through the batch can be also carried out using the slider.

- Figure 4 (B) shows the set of annotation utilities. To ensure sufficient functionalities for the annotator, from that module, one can choose between the buttons to draw freehand, annotate with dots, save annotations to the database, or delete annotations.

- Figure 4 (C) WSI labeling tool. This module allows the pathologist to set the global label of the WSI with

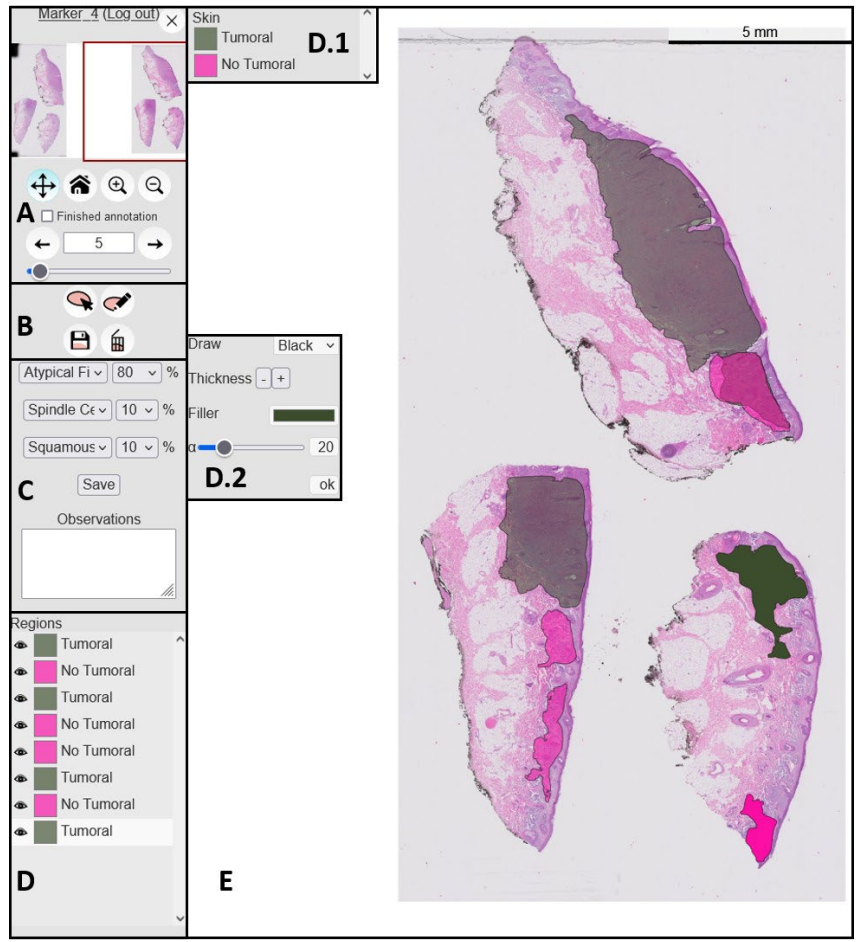

**Figure 4:** Main functionalities of the proposed web application. **A** Visualization and navigation. **B** Annotation tools. **C** WSI label and observation. **D** Annotations and features. **E** WSI viewer.

a certain degree of uncertainty; in this example use case, the annotator can choose among three different labels to annotate the WSI and introduce the degree of certainty in the annotation. In addition, the annotator can write comments about the diagnosis in the text Observations box.

- Figure 4 (D) Region annotations. In this area, annotations are indexed with three elements: an eye-shaped icon that allows to hide or show the annotations, a square with the color of the drawn region, and a text with the label name. Clicking on each annotation brings up two submenus:

    – Figure 4 D.1) A labeling panel used to choose the label of the selected annotation.
    – Figure 4 D.2) shows a window to configure the visual aspects of the annotation: region line color, thickness, background color and transparency.

- Figure 4(E) WSI viewer. This is the canvas on which the WSI being analyzed is displayed. On this canvas, the ROIs are visualized by drawing them in color according to the applied label.

In the example use case, using the tools described above, two expert dermatopathologists analyzed all slides and assigned a global label per WSI. The pathologists annotated the tumor regions of some slides belonging to the database. In this sense, the WSIs were divided into regions of interest or ROIs (tumor regions) and non-interest (the rest of the WSI).

### 3.1.2. Heat-map prediction

Once the model developed for a concrete application predicts the interest regions (tumor regions in our use case example) at the patch level, it is necessary to obtain the tumor region of the entire WSI. To do that, we build a heat-map for each WSI; see Figure 1 (C). In probability heat maps of each pixel, the tumor region probabilities predicted are estimated by bilinearly interpolating the predicted probabilities of the closest patches in terms of Euclidean distance to the center of the patches. Each heat map is converted into a grey scale image, and thresholded obtaining a binary mask. Subsequently, the mask is filtered, and the contours of the predicted tumor regions are smoothed.

The absolute coordinates of the tumor region located in the WSI are inserted into the web server database by performing an SQL insert query and specifying the following fields: batch name, user name, slide name, and the coordinates, see Figure 2 (C). This process transforms the WSI tumor regions predictions in annotations, allowing them to be displayed in the web service.

### 3.2. Validation

In order to demonstrate and validate the tool presented in this paper, this section will describe the selected use case: a model for segmenting regions of interest for spindle cell skin neoplasms.

#### 3.2.1. Model training

The aim is to build, from the small number of annotations made by the experts, a model able to automatically predict the tumour regions for any image that is uploaded to the system, see Figure 1 (B).

**Patch extraction.** To process the large WSIs by a deep learning model, these were downsampled to 10x resolution and divided into patches of size 512x512x3 with a 50 % overlap. Aiming at pre-processing the biopsies and reducing the noisy patches, a mask indicating the presence of tissue in the patches was obtained by applying the Otsu method over the magenta channel. Patches with less than 20% of tissue were excluded from the training.

**Model optimization.** To optimize the classifier, we use the annotated and non-annotated images. Formally, we denote $D = \{X, Y\}$ as the independent training set, where $x_i \in X$ refers to the $i$-patch of the annotated dataset with its corresponding ground-truth $y_i \in Y$, being $i = 1, 2, ..., N$ the number of training patches pair. Additionally, we define the unlabelled set as $D^T = \{X^T\}$. Our proposed approach uses a teacher-model paradigm to improve the selection of tumoral regions at the patch level.

*Teacher model.* The teacher model aims to classify high-confidence patches using pathologists' labels for learning. Therefore, given a histological image $x \in \mathbb{R}^{M \times N \times d}$, where $M \times N \times d = 224 \times 224 \times 3$, a feature-embedded map $z = f_\varphi(x)$ with $z \in \mathbb{R}^{H \times W \times C}$, is provided by the feature extractor $f_\varphi$. After that, the embedding representations are fed into a classification layer ($g_\theta$) to obtain class probabilities of labeled data sets using a Softmax function, $\hat{Y}^T$.

*Student model.* The student model aims to improve the prediction of the teacher model based on a pseudo-supervised and supervised dataset of images, using the teacher model predictions as pseudo-labels. First, all instances from the unlabelled dataset are predicted using the teacher model. Last, the pseudo-labels are used to augment the training dataset, which results in $D^S = \{X \cup X^T, Y \cup \hat{Y}^T\}$. In this way, the student model in the last step is trained on the augmented pseudolabeled dataset. For training the proposed model, we weighted the cross-entropy loss function:

$$L(y, \hat{y}) = \sum_i \omega_i^T y_i \log(\hat{y}_i) \quad (1)$$

**Table 3**
Hospital, year of residence and images annotated by each non-expert pathologist involved in the validation process.

| | Hospital | Residence (year) | Total Images |
|---|---|---|---|
| Non-expert 1 | Granada | R4 | 228 |
| Non-expert 2 | Granada | R3-R4 | 233 |
| Non-expert 3 | Granada | R3 | 229 |
| Non-expert 4 | Granada | R4 | 226 |
| Non-expert 5 | Granada | R3 | 233 |
| Non-expert 6 | Granada | R3-R4 | 227 |
| Non-expert 7 | Valencia | R1 | 237 |
| Non-expert 8 | Valencia | R3 | 227 |
| Non-expert 9 | Valencia | R1 | 231 |
| Non-expert 10 | Valencia | R2 | 234 |

where

$$\omega_i^T = \begin{cases} 1, & \text{if } x_i \in X \\ \max\left(\sum \frac{\exp(g_{\theta}(f_{\varphi}(x)))}{\sum_c \exp(g_{\theta}(f_{\varphi}(x_c)))}\right) & \text{if } x_i \in X^T \end{cases} \quad (2)$$

$\omega_i^T$ introduces the confidence of the teacher model to the student model, improving the prediction.

Therefore, by optimising the classifier of proposed use-case, it is possible to obtain the classification of all the patches composing the WSIs in the database into tumour or non-tumour.

#### 3.2.2. Pathologists-in-training framework

As mentioned above, to validate the HistoColAi software in a real scenario, ten non-expert pathologists participated in the study. In this case, four resident pathologists belonged to the University of Valencia and six to Granada. An annotation protocol was designed to ensure that 106 WSIs were annotated by all non-expert annotators (dense set). In contrast, the rest were only annotated by some pathologists (non-dense set). Table 3 shows the summary of the WSI distribution per annotator.

The annotation protocol designed for non-expert pathologists included two annotation levels (WSI and pixel), see Figure 5. Note that an image batch was created for each annotator. In this way, each pathologist-in-training can only see their own annotations.

**Pixel level**: The tumor region predicted by the *Histo-ColAi* classifier was the input for the training pathologists' workflow. In this vein, each non-expert pathologist assigned a tumor or non-tumor label to each region automatically predicted by the deep learning system. It was enforced that for an annotator to consider a region as "tumoral", at least 50% of the region content must be tumoral. In addition, each pathologist had the draw button enabled to annotate other tumor regions considered as tumoral that were not automatically selected. It is worth mentioning that, the non-expert pathologists used the automatic predictions of the model for each image uploaded to the system to enhance and speed up the selection of tumor regions.

**WSI level**: Each non-expert pathologist assigned a global label (image-level) to each WSI corresponding to one of the seven considered types of neoplasms.

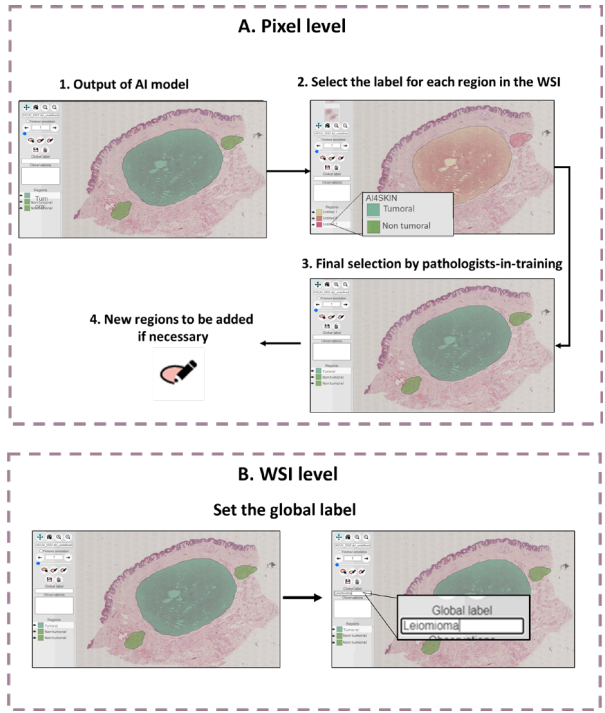

**Figure 5:** Workflow for non-expert pathologists at pixel and WSI level using the HistoColAi with the prediction of the interest region model.

**Table 4**
Evaluation of the tumor region model in the subset of images annotated by experts.

|         | ACC    | F1-score | AUC    |
|---------|--------|----------|--------|
| lm      | 0.9420 | 0.9600   | 0.9501 |
| lms     | 0.8677 | 0.9200   | 0.8169 |
| df      | 0.8897 | 0.9266   | 0.8345 |
| dfs     | 0.8105 | 0.8486   | 0.8029 |
| mfc     | 0.8135 | 0.8606   | 0.7800 |
| fxa     | 0.8840 | 0.9091   | 0.8973 |
| cef     | 0.9147 | 0.9191   | 0.9141 |
| Average | 0.8746 | 0.9063   | 0.8600 |

## 4. Results and discusion

In this section, we analyze the model performance for the tumor region prediction based on the annotation of two expert pathologists. Additionally, we evaluate the app usability in the pathologists-in-training scenario.

### 4.1. Analysis of the model performance

**Model vs. expert**. To evaluate the performance of the tumor region model, we used the subset of images annotated by experts (15% of the database). In this case, we evaluate the model at patch level. Table 4 shows the model's performance using the validation set annotated by the expert pathologist. To this end, different figures of merit, e.g., accuracy (ACC), F1-score (FS), and area under the ROC curve (AUC), are considered.

**Table 5**
Analysis of the model performance against the decisions of non-experts in the dense set. Note that the mean and standard deviation of non-experts is calculated for each type of neoplasm.

|         | $Recall_T$          | $Recall_{NT}$       |
|---------|---------------------|---------------------|
| lm      | 0.8700 ± 0.0692     | 0.9205 ± 0.0776     |
| lms     | 0.9406 ± 0.0516     | 0.9552 ± 0.0471     |
| df      | 0.9196 ± 0.0484     | 0.9794 ± 0.0360     |
| dfs     | 0.9562 ± 0.1104     | 0.9189 ± 0.1499     |
| mfc     | 0.8927 ± 0.0632     | 0.8883 ± 0.1040     |
| fxa     | 0.9115 ± 0.0652     | 0.9683 ± 0.0555     |
| cef     | 0.8558 ± 0.0885     | 0.9471 ± 0.0936     |
| Average | 0.9066 ± 0.0363     | 0.9397 ± 0.0320     |

As shown in Table 4 the performance of the model is high for most of the neoplasms under study. Dermatofibrosarcoma (dfs) and spindle cell melanoma (mfc) have the lowest metrics. Dermatofibrosarcoma tumor cells do not show atypia (a common feature of tumor cells), which explains why the model can misidentify tumor regions with normal fibroblasts (non-tumoral region). In the case of spindle cell melanomas, the non-tumor tissue has similar features to scar tissue, which would clarify why the model confuses them with tumor regions.

**Model vs. non-experts**. This section presents the evaluation of the model performance against decisions carried out by non-expert pathologists on the automatically predicted regions. In this case, the assessment is performed at region level. Note that, for each slide, non-expert pathologists automatically obtain the model predictions of the tumor regions. At this stage, they should consider whether the proposed regions are tumor or non-tumor and annotate the possible tumor regions missed by the model. To evaluate the success rate of model vs. non-experts, we propose the following metrics:

$$Recall_T = \frac{TM - NT}{TM} \quad (3)$$

$$Recall_{NT} = \frac{TM - NT}{TP} \quad (4)$$

where TM represents the tumor regions detected by the model, NT is the tumor regions detected by the model that the non-experts consider as non-tumor, and TP is all the regions that the non-expert pathologists consider as tumor. Note that TP regions may be larger than TM regions as the non-expert pathologist can annotate new regions considered tumor. $Recall_T$ measures the model performance by identifying tumor regions, while $Recall_{NT}$ identifies the model goodness in the identification of non-tumor regions. Table 5 shows the results obtained. Only images of the dense set (analyzed by all non-expert pathologists) were considered.

The $Recall_T$ and $Recall_{NT}$ metrics are high for all the neoplasms under study, reinforcing that the developed model strongly predicts tumor regions. In most cases, the $Recall_{NT}$ metric is higher than $Recall_T$, which would explain the tendency of the model to detect all tumor regions within the WSI, even if some of them are considered by non-experts as non-tumor regions. As in Section 4, mfc neoplasm had the worst performance. In addition, the most significant discordance among non-expert pathologists is found in dfs neoplasm, as the standard deviation is higher. This fact demonstrates the correlation between the decisions taken by non-experts and the annotations made by experts.

### 4.2. Usability analysis

This section describes the usability analysis carried out in the use-case using HistoColAi.

**Non-expert's background**. Before the non-experts started annotating with the developed application, a survey was conducted asking them which annotation tools they knew or used. Of the ten non-expert pathologists, six had not previously used any digital segmentation tool, two were familiar with Qupath (Non-expert 2 and Non-expert 10), one with ImageJ (Non-expert 3) and one with Cell IA (Non-expert 1). The two users who had used Qupath found the tool difficult to use because it was not intuitive. In contrast, users who had worked with Cell IA and Image J found them easy to use. The survey results indicate a limited widespread use of software for WSI image annotation. This fact can be explained by the difficulty of access, as most of them need to be installed, and those that are web accessible are private platforms (Upath), and non-ad-hoc services (ImageJ, Table 1). Therefore, the results demonstrate the lack of an easy-to-use ad-hoc tool for pathologists' annotation tasks.

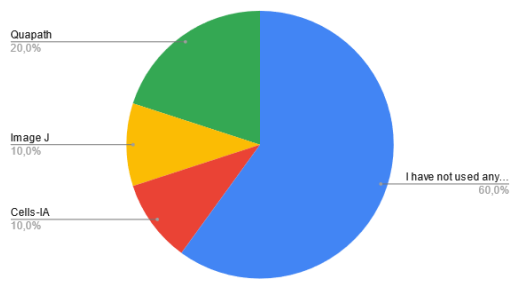

**Figure 6:** Non-experts survey of results knowledge in annotations tools.

**HistoColAi impact on the performance of non-experts in identifying the class of each neoplasm**. To perform this analysis, the database was divided into two batches spaced one month in time. Table 6 shows the average of the ACC and F1-score metrics obtained by each non-expert pathologist identifying the neoplasm in the two consecutive batches. The table shows that ACC and F1-score improve in 80

In addition, the developed tool allows assessing the performance of pathologists-in-training in detecting the different types of neoplasms under study. Figure 7 shows a Venn diagram highlighting the discrepancies between the expert annotator and the in-training annotators. For each type of neoplasm, the diagram is composed of two overlapping circles. The red circle represents the number of WSIs labeled by the expert and the green one by the non-experts. The overlapping area indicates the number of coincidences between them. We can state that non-experts generally perform the best in labeling benign lesions, i.e. leiomyoma and dermatofibroma. For most malignant neoplasms, leiomyosarcoma, dematofibrosarcoma, fibroxanthoma and especially carcinoma, the agreement between experts and non-experts is low. Figure 8 shows the confusion matrix of each non-expert pathologist. Pathologists in training are confused by the classes from the same tissue, leiomyoma vs. leiomyosarcoma and dermatofibroma vs. dermatofibrosarcoma. In both cases, non-experts tend to label regions with benign labels (leiomyoma or dermatofibroma). Analyzing the data disaggregated by a non-expert pathologist, we can see that the non-experts who perform better on difficult labels are those with more years of residency.

**Table 6**
Comparison of the performance of non-expert pathologists between two consecutive batches measured in accuracy (ACC) and F1-score. Note that, in this case, both the dense and the non-dense set were included.

|  | ACC | | FI-score | |
|---|---|---|---|---|
|  | Batch 1 | Batch 2 | Batch 1 | Batch 2 |
| **Non-expert 1** | **0.6989** | 0.5851 | **0.5500** | 0.4709 |
| **Non-expert 2** | 0.5121 | **0.6363** | 0.41389 | **0.5197** |
| **Non-expert 3** | 0.4750 | **0.4800** | 0.3602 | **0.3980** |
| **Non-expert 4** | 0.5984 | **0.7100** | 0.4900 | **0.5360** |
| **Non-expert 5** | 0.5652 | **0.6408** | 0.4700 | **0.5205** |
| **Non-expert 6** | **0.6907** | 0.5703 | **0.5410** | 0.4831 |
| **Non-expert 7** | 0.5400 | **0.5960** | 0.4000 | **0.4634** |
| **Non-expert 8** | 0.5333 | **0.5878** | 0.4174 | **0.4700** |
| **Non-expert 9** | 0.4123 | **0.4300** | **0.3180** | 0.3100 |
| **Non-expert 10** | 0.6000 | **0.6131** | 0.4698 | **0.5000** |

Finally, HistoColAi allow us to monitor the time taken by each pathologist in the WSI analysis. Figure 9 shows the Ribbon chart, which gives information on the average time (in minutes) spent by each non-expert pathologist to label each type of neoplasm. The non-expert pathologists who spent the least time on each slide were 9 and 10. This fact, combined with the years of experience (first and second year of residency, respectively), results in a low level of performance. The non-experts with the best performance results (1, 4 and 6, in the fourth and third year of residency) had similar average times. However, they show differences in the annotation time per class. Note that non-expert 1 had previous experience with digital annotation tools. The fact that this pathologist takes a similar time to non-experts 4 and 6 demonstrates that our tool is easy to use and does not require previous experience. In the case of the most experienced non-experts (3-4 years of residency), the types of neoplasm that took the longest to be labeled were Dermatofibrosarcoma (DFS), Fibroxanthoma (FXA), Leiomyosarcoma (LMS) and Carcinoma (CEF), which correspond to malignant cases. In general, labeling a WSI took 2 minutes and 49 seconds. Therefore, the use of the proposed

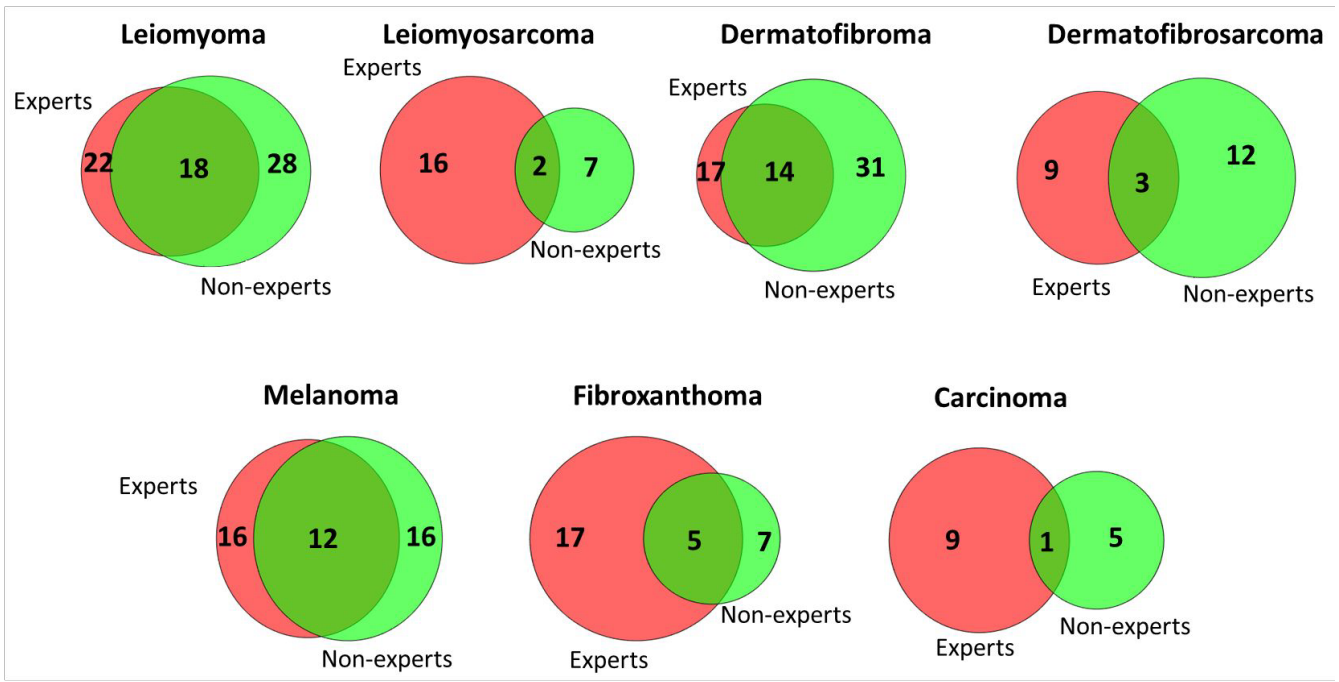

**Figure 7:** Venn diagrams illustrating the overlap in WSI identification produced by expert pathologists and non-expert pathologists for each type of neoplasm. Note that in the case of non-experts the the number corresponds to the average number of WSIs annotated by the non-experts with the corresponding label.

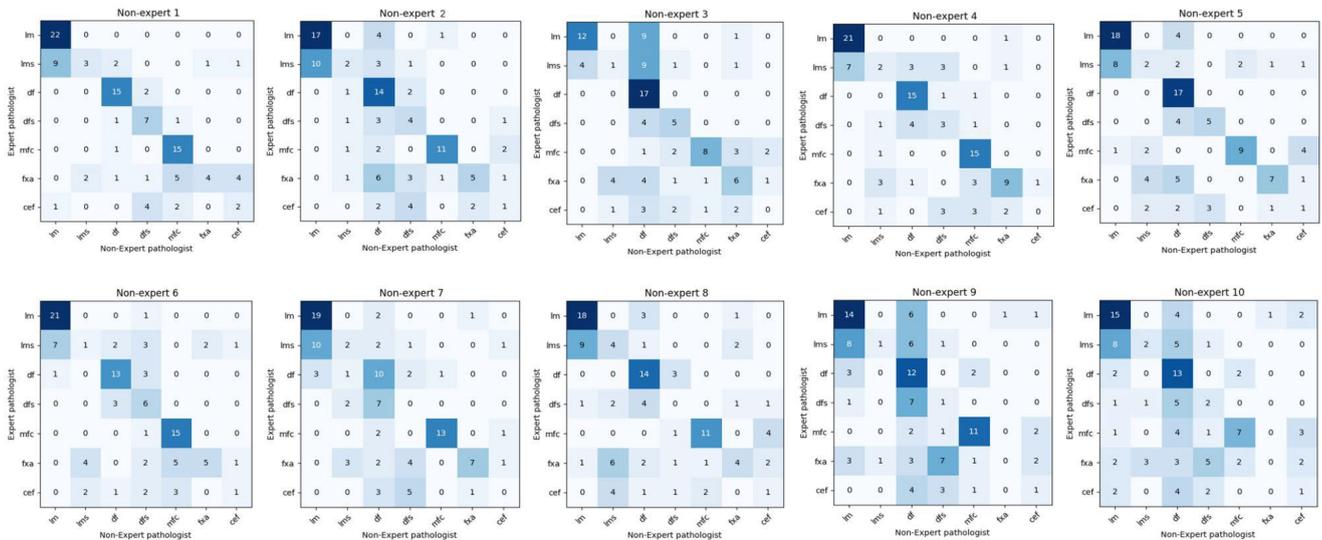

**Figure 8:** Inter-pathologist variability for identifying the seven neoplasms under study. All confusion matrices were normalized per row, reflecting each class's recall metric.

application significantly decreases the time a pathologist must spend labeling an image.

## 5. Conclusion

In this paper, we propose a new web interface app for WSI annotation and classification. The proposed tool development is a breakthrough in the digital pathologist field. *HistoColAi* is one of the first open-source web services with embedded artificial Intelligence models for the automatic detection of tumor regions in WSIs. In addition, the developed tool allows performance studies by analyzing the time required by each user for the annotation of the WSIs. To perform a usability test, the application was validated under a pathologist-in-training paradigm. Using the application in a pathologist workflow will create a sizeable heterogeneous dataset to train accurate deep learning models. Therefore, this application contributes significantly to advancing the streamlining of digital diagnostics.

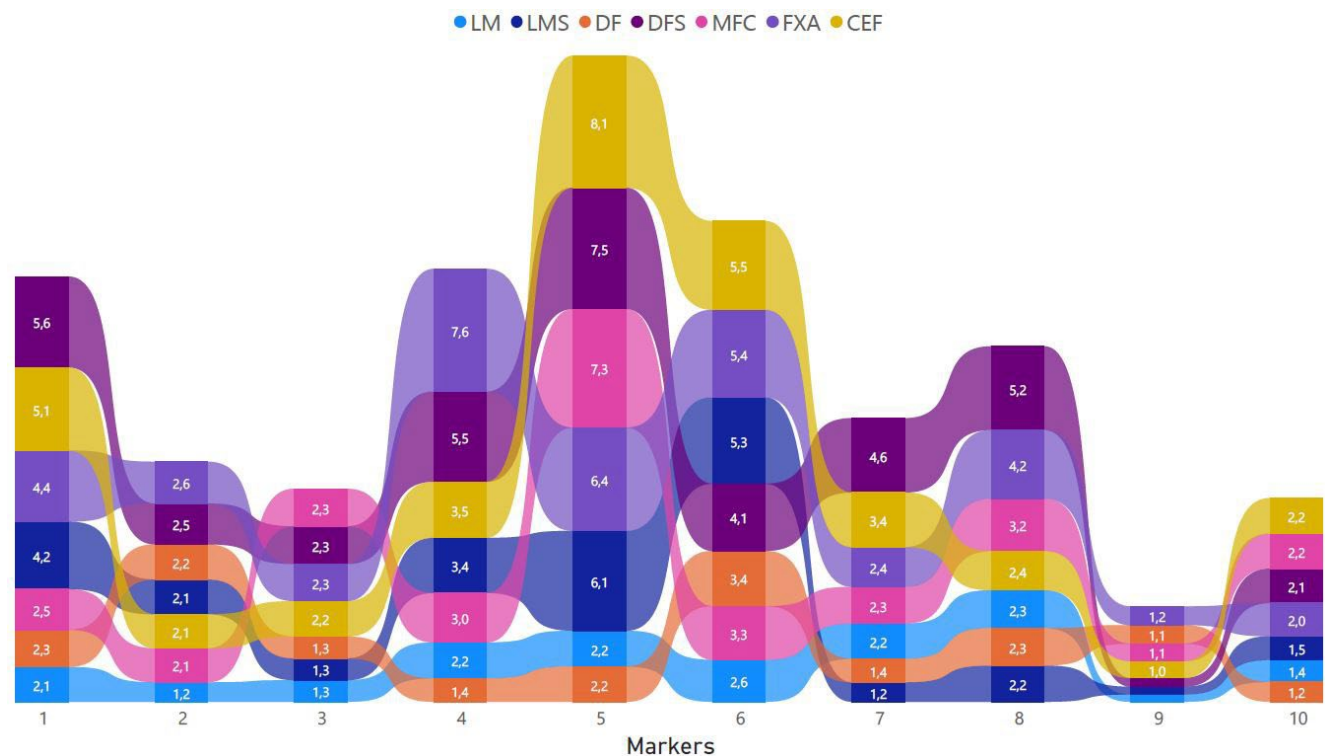

**Figure 9:** Ribbon chart with the average non-experts labeling time by type of neoplasm. Note that this graph shows the average time per WSI in minutes for the annotation and classification task.

## Acknowledgements

We gratefully acknowledge the support from the Generalitat Valenciana (GVA) with the donation of the DGX A100 used for this work, action co-financed by the European Union through the Operational Program of the European Regional Development Fund of the Comunitat Valenciana 2014-2020 (IDIFEDER/2020/030).

## Funding

This work has received funding from Horizon 2020, the European Union's Framework Programme for Research and Innovation, under the grant agreement No. 860627 (CLARIFY), the Spanish Ministry of Economy and Competitiveness through project PID2019-105142RB-C21 (AI4SKIN) and GVA through the project INNEST/ 2021/321 (SAMUEL). he work of Cristian Camilo Pulgarín Ospina has been supported by the Spanish State Research Agency (PRE2020-093271). The work of Rocío del Amor has been supported by the Spanish Ministry of Universities (FPU20/05263). The work of J. Silva-Rodríguez was carried out during his previous position at Universitat Politècnica de València.

## Conflict of interest

The authors declare that they have no conflict of interest.